\documentclass[aps,pra,superscriptaddress,twocolumn,amsmath,amsfonts,amssymb,floatfix,nofootinbib]{revtex4}
\usepackage{enumerate}
\usepackage{mathtools}
\usepackage{amsmath}
\usepackage{graphicx}
\usepackage{epsfig}
\usepackage{sidecap}
\usepackage{hyperref}
\usepackage[normalem]{ulem}
\usepackage{color}
\usepackage{enumerate}
\usepackage{bm}

\usepackage[utf8]{inputenc}
\usepackage{amsmath}
\usepackage{amsfonts}
\usepackage{amssymb}
\usepackage{graphicx}
\usepackage{enumitem}

\begin{document}

\title{Comparing coherence measures for $X$ states: Can quantum states be ordered based on quantum coherence? }

\author{Sandeep Mishra} \thanks{sandeep.mtec@gmail.com }
\affiliation{University School of Basic and Applied Sciences, Guru Gobind Singh Indraprastha University, Sector 16 C, Dwarka, New Delhi-110078, India}

\author{Kishore Thapliyal} \thanks{tkishore36@yahoo.com}
\affiliation{Jaypee Institute of Information Technology, A-10, Sector-62, Noida, UP-201309, India}
\affiliation{RCPTM, Joint Laboratory of Optics of Palacky University
and Institute of Physics of Academy of Science of the Czech Republic,
Faculty of Science, Palacky University, 17. listopadu 12, 771 46 Olomouc,
Czech Republic}

\author{Anirban Pathak} \thanks{anirban.pathak@gmail.com}
\affiliation{Jaypee Institute of Information Technology, A-10, Sector-62, Noida, UP-201309, India}
   
\author{Anu Venugopalan}\thanks{anu.venugopalan@gmail.com}
\affiliation{University School of Basic and Applied Sciences, Guru Gobind Singh Indraprastha University, Sector 16 C, Dwarka, New Delhi-110078, India}

\begin{abstract}
Quantum coherence is an essential resource for quantum information processing and various quantitative measures of it have  been introduced. However, the interconnections between these measures are not yet understood properly. Here, using a large set of randomly prepared $X$ states and analytically obtained expressions of various measures of coherence (e.g., relative entropy of coherence, $l1$ norm of coherence, coherence via skew information, and first-order coherence), it is established that these measures of quantum coherence cannot be used to perform ordering of a set of quantum states based on the amount of coherence present in a state. Further, it is shown that  for a given value of quantum coherence measured by the relative entropy of coherence,  maximally nonlocal mixed states of $X$ type (which are characterized by maximal violation of the CHSH inequality) have maximum quantum coherence as measured by $l1$ norm of coherence. In addition,  the amount of coherence measured by $l1$ norm of coherence for a Werner state is found to be always less than that for a maximally nonlocal mixed state even when they possess an equal amount of coherence measured by the relative entropy of coherence. These resource theory based measures of coherence are not observed to show any relation with the first-order coherence, while its maximum (hidden coherence) is found to be more connected to concurrence both being basis independent quantities. These observations could be of use in obtaining a deeper understanding of the interconnections between various measures of quantum coherence.
\end{abstract}

\maketitle

\section{Introduction} \label{Intro}
{ \textit {Coherence}} is ubiquitous in quantum systems and is known to be the primary factor behind the emergence of phenomena that are dramatically different from those observed in the systems where it is absent.  This fact was first realized in the context of wave optics, where the  presence of coherence was observed to have some effects which were radically different from the  phenomena usually observed in the domain of traditional geometrical optics \cite{Mandel,Boyd,cls}. More dramatic consequences of coherence, specifically that of quantum coherence which signifies the existence of superposition between quantum states, appeared in the last century with the advent of quantum mechanics in general and with the understanding of quantum interference in particular. The interest in quantum coherence further amplified with the advent of quantum computation and communication, where entangled states, which are nothing but non-separable superpositions in a tensor product space,  play a crucial role. 

Quantum coherence originates from the wave function description of quantum systems and cannot be described within the framework of classical physics.  A consequence of quantum coherence is the existence of nonclassical states which have no classical analogue and can only be understood as uniquely quantum in character \cite{glaub,sudar}. These states are essential for the establishment of quantum  supremacy (\cite{sup} and references therein). In fact, in the context of quantum information processing,  quantum coherence {is being widely accepted as an important}  resource \cite{coh_rev2,coh_rev1}, and thus it is important to quantify the amount of  coherence present in a quantum state. {Recently,} Baumgratz et al. \cite{plenio} have provided a framework for the  quantitative characterization of quantum coherence by {treating it} as a physical resource. The framework to quantify coherence is based on considering an incoherent basis and defining an incoherent state as one which is diagonal in that basis.  Since the pioneering work of Baumgratz et al. \cite{plenio}, various measures of quantum coherence have been proposed \cite{str, rana, qir, yu, roc, yuan, winter, chitambar} and used \cite{coh_rev2,coh_rev1}.  However, till date only the relative entropy of coherence, {$l1$ norm} of coherence, and a skew information based measure of coherence have been found to be satisfactory. It is worth noting here that all these measures are basis dependent quantities. In what follows, we will discuss these measures along with some other proposed measures of quantum coherence as well as a measure of  first-order coherence \cite{optical_coh}, which can also be referred to as optical coherence as it resembles the measure of coherence used in optics. 

Apart from the quantitative measures of coherence, the idea of  quantum coherence has also been examined from various other interesting perspectives.  For example, in the field of quantum thermodynamics, studies have been performed to understand the possibilities of the extraction of work from quantum coherence \cite{szilard}.  Studies of low temperature thermodynamics with quantum coherence \cite{nat1,nat2} and the role of quantum coherence in energy transfer \cite{pre} have also been reported. In  quantum biology \cite{bio1}, it has been shown that quantum coherence plays a major role in photosynthesis \cite{bio2}, and quantum coherence and entanglement are  known to play an important role in the avian compass {in} migratory birds \cite{bio3}. Similarly, quantum coherence has various important applications in the field of quantum algorithms \cite{alg1, alg2} and quantum metrology \cite{met1,met2}.  

Further, the close {relationship} among nonclassicality, entanglement, Bell nonlocality, and quantum coherence {leads} to {the} question: are the {known} limitations of the quantitative measures of nonclassicality  \cite{adam,adam-non}, entanglement \cite{adam-ent}, steering \cite{adam-st}, and Bell nonlocality \cite{adam-bi}  also apply to the quantitative measures of quantum coherence? In all earlier studies \cite{adam,adam-non,adam-ent,adam-st,adam-bi,adam-nm}, it has been observed that the ordering of quantum states based on the amount of a particular type of nonclassicality they contain (as measured by different quantitative measures of that nonclassical feature)  is usually inconsistent.  Therefore, here we attempt to answer a question: If a measure of coherence {$p$,} indicates that state {$A$} has more coherence than state {$B$}, will that {also} mean that {$A$} will always  be found to have more coherence even when coherence {is measured} using another quantitative measure, {$q$,} in the same basis? In other words, are the quantitative measures of the quantum coherence monotone of each other? If not, are they connected with each other for {a subset} of a family of quantum states? Until now, no effort has been made to answer these questions satisfactorily. Motivated by these facts, here we aim to address some of the above mentioned hitherto unanswered questions related to the measures of quantum coherence and their interrelations using $X$ states \cite{xstate1, xstate2} as our example states as these states are known to serve as a good test bed to {explore and} study the properties and applications of quantum coherence \cite{x_dynamics,xh,xp_p1, xp_p2, xp_p3, xp_p4, xp_uc1, xp_uc2, xp_nmr}. In what follows, we will show that the quantitative measures of quantum coherence are not monotone of each other unless  they are trivially dependent on each other, but are only related for a subset of a family of $X$ states, and such states define the boundary values of the measures of coherence.  By obtaining the `relative' coherence (the amount of coherence using one measure relative to that from another measure) we could identify the states having maximum and minimum values of relative coherence. This in turn can help us to illuminate the connections between different measures of quantum coherence.

The rest of the paper is organized as follows. In Section \ref{measures}, we briefly review different measures of coherence. Thereafter, we introduce $X$ states and their properties in Section \ref{X states}. This is followed by our comparative study in Section \ref{our-res} to establish the relationship between different measures of coherence for the family of $X$ states. We briefly discuss first-order coherence for $X$ states in Section \ref{Op-coh} before concluding the paper in Section \ref{con}. 

\section{Measures of coherence}\label{measures}
As mentioned above, Baumgratz et al. \cite{plenio} have recently provided a prescription for the quantitative characterization of quantum coherence by considering coherence as a physical resource. They proposed the following set of criteria (which we would refer to as Baumgratz et al.'s criteria) that every potential quantifier of coherence $(C)$ should satisfy:
\begin{itemize} 
\item $(C1)$ Non-negativity:  $C(\rho)\geq 0$ with equality if and only if $\rho$ is incoherent.
\item $(C2a)$ Monotonicity: $C$ does not increase under the application of completely positive and trace preserving incoherent operations, i.e., $C(\phi[\rho])\leq C(\rho)$, where $\phi$ is any completely positive and trace preserving incoherent operation.
\item $(C2b)$ Strong monotonicity: $C$ does not increase on an average under selective incoherent operations, i.e., $\underset{i}{\sum} q_{i} C(\rho_{i})\leq C(\rho)$, where $\rho_{i}=(K_{i}\rho K_{i}^{\dagger})/q_{i}$ are post measurement states with probabilities given by $ q_{i}= Tr[K_{i}\rho K_{i}^{\dagger}] $, and $K_{i}$ are incoherent Kraus operators.
\item $(C3)$ Convexity: $C$ does not increase under mixing, i.e., $ \sum_{i} p_{i} C(\rho_{i}) \geq C(\sum_{i}  p_{i}\rho_{i})$.
\end{itemize}

Here, $\rho$ is the density operator corresponding to the quantum state. Based on the above criteria, many quantum coherence measures have been introduced. However, it is very difficult to prove the condition of strong monotonicty $(C2b)$ for them. Usually, it is sufficient to just prove the conditions $(C1)$, $(C2b)$ and $(C3)$ as $(C2a)$ is already implied by $(C2b)$ and $(C3)$. It is found that of all the existing well-known measures of coherence, only the relative entropy of coherence and $l1$ norm of coherence satisfy these criteria and hence these two measures of coherence serve as good quantifiers of coherence. Along with these, a recently introduced coherence measure based on skew information has also been shown to satisfy Baumgratz et al.'s criteria \cite{yu}. In this section, we aim to briefly describe these three measures along with some other proposed quantifiers with an aim to compare them and  find out their interrelations and limitations. To begin with, we describe relative entropy of coherence.

\subsection{Relative entropy of coherence}

The relative entropy of coherence \cite{plenio} present in a quantum state represented by the density matrix $\rho$ is defined as 
\begin{equation}\label{rel_ent}
C_{\rm{} rel}(\rho)= S(\rho_{\rm{} diag})- S(\rho),
\end{equation}
where $S(\rho )$ is the von Neumann entropy of $\rho$, and $\rho_{\rm{} diag} $ denotes the state obtained from $\rho $ by removing all the off-diagonal elements of $\rho$. Note that $C_{\rm{} rel}$ (\ref{rel_ent}) is a basis dependent quantity. Due to its similarity in form to that of the relative entropy of entanglement, $C_{\rm{} rel}$  has a physical meaning \cite{winter}. Specifically, it physically represents the optimal rate of the distilled maximally coherent states that can be produced by incoherent operations in the asymptotic limit of many copies of $\rho$. Interestingly, experimental measurement of this coherence quantifier can be performed without full quantum state tomography \cite{yu}.

\subsection{$l1$ norm of coherence}

The $l1$ norm of coherence \cite{plenio} is given by
\begin{equation}\label{l1}
C_{l1}(\rho)= \sum _{i,j,i\neq j}|\rho_{i,j}|.
\end{equation}
This measure of coherence, which like $C_{\rm{} rel}$ (\ref{rel_ent}) is also basis dependent, is presently not known to have any analogue in the resource theory of entanglement \cite{coh_rev2}.  While efforts had been made earlier to find a physical interpretation to (\ref{l1}), recently it has been reported that for a multi-slit interference set up coherence as defined by (\ref{l1}) can be experimentally measured \cite{bera,bagan, tania, biswas, sandeep}. 

\subsection{Skew information based measure of coherence}

In 2014, Girolami \cite{qir} introduced a new experimentally accessible measure of  coherence known as $k-$coherence which was based on quantum skew information. The motivation for this was the fact that quantum coherence of a state $\rho$ is rooted in unpredictability, and the state $\rho$ is incoherent in the eigenbasis of an observable  $k$ if and only if it commutes with $k$. The $k-$coherence is given by:
\begin{equation}
I(\rho, k)= - \frac{1}{2} Tr \{[\sqrt{\rho},k ]^{2}\} \label{K}
\end{equation}
Later it was found that  (\ref{K}) violates the property of strong monotonicity for certain states \cite{qir_not}. Recently,  motivated by the $k-$coherence, Yu \cite{yu} has proposed a new quantum coherence measure using skew information and have proven that it satisfies all the conditions required for a good quantifier (mentioned above as Baumgratz et al.'s criteria).  Additionally, in contrast to a set of other measures of coherence mentioned in the following subsection, it has an analytic expression which is easy to calculate and analyze. The quantum coherence of state $\rho $ in the basis $\{ | i \rangle \}$ via skew information is given by
\begin{equation}\label{skew}
C_{\rm{} skew}(\rho)= - \frac{1}{2}\sum _{i}Tr \{[\sqrt{\rho},| i \rangle \langle i |]^{2} \}.
\end{equation}

It is important to note that all the above measures of coherence (i.e., relative entropy of coherence, $l1$ norm of coherence, and skew information based coherence) satisfy Baumgratz et al.'s criteria and have closed form expressions, hence their computation does not involve any optimization method. Further, relative entropy of coherence and $l1$ norm of coherence are considered to be equally good measures of quantum coherence, and  a relation between these two measures has been conjectured by Rana et al. \cite{rana} as 
\begin{equation}\label{conj}
C_{l1}(\rho)\geq C_{\rm{} rel}(\rho).
\end{equation}
{Rana et al.}  have proved the conjecture (\ref{conj}) for pure qubit states, but the validity of this conjecture for mixed states is still an open problem. Further, no such relation between $l1$ norm (or relative entropy) based measures of coherence and skew information based measure of coherence has yet been investigated. In what follows, we will try to provide some insights into these issues using the example of the class of $X$ states.

\subsection{Other quantitative measures of quantum coherence}

Nonclassicality measures have traditionally been studied with reference to the quantum theory of light (see Ref. \cite{adam} and references therein). Intuitively, it seems obvious that nonclassicality in light must be connected with the notion of  quantum coherence. This is so because a nonclassical state is defined as a state which cannot be expressed as a mixture of coherent states. This essentially implies the existence of non-zero coherence (off-diagonal terms in the density matrix of a nonclassical state) when  expressed in the coherent state basis. Naturally, a set of  quantifiers of quantum coherence analogous to earlier proposed measures of nonclassicality have been proposed. For example, in the context of nonclassicality, a measure of nonclassicality in single mode fields was introduced in the past \cite{asb} as the amount of two mode entanglement generated by it (i.e., a nonclassical state) using linear optics and classical states only. In analogy, Streltsov et al. \cite{str} have tried to relate the theories of quantum coherence with that of quantum entanglement. They have proved that any degree of quantum coherence with respect to some reference basis can be converted to entanglement via application of incoherent operations. This approach is similar to Asboth et al.'s idea \cite{asb}, which provided a model for inter-convertibility of the quantifiers of nonclassicality and entanglement.

To understand the idea of Streltsov et al. \cite{str}, let us consider an example with a source (S) (which may or may not have coherence) attached to an ancilla (A) and apply a $CNOT_{SA}$ operation jointly to the source and ancilla. If the ancilla and source are in the states $|0\rangle_{A}$ and $|0\rangle_{S}$, respectively, the application of $CNOT_{SA}$ gate on the joint state of source and ancilla would result in the state $|00\rangle_{SA}\rightarrow |00\rangle_{SA}$, which is clearly a separable state. When the state of source is $\alpha|0\rangle+\beta|1\rangle: |\alpha|\neq0 $ and $|\alpha|^2+|\beta|^2=1$, i.e., a state with a finite amount of coherence,  the application of the $CNOT_{SA}$ gate on the joint state of source and ancilla would entangle them, i.e., $\alpha|0\rangle_{S}+\beta|1\rangle_{S} \otimes |0\rangle_{A}\rightarrow \alpha|00\rangle_{SA}+\beta|11\rangle_{SA} $. Clearly, $CNOT_{SA}$ gate is acting here as an incoherent operator and producing entanglement between the source and the ancilla only if the source has some quantum coherence in it. Further, upon the application of $CNOT_{SA}$ gate, the amount of entanglement created between the source and the ancilla comes at the cost of amount of coherence left over in the source state. 

It is not our purpose to elaborate on this analogy. However, this analogy hints at the possibility that the limitations of nonclassicality measures may be present in the coherence measures, too. As we will further discuss that some measures of quantum coherence require optimization over infinitely many states and are expected to encounter the same drawbacks as similar nonclassicality measures are known to face \cite{adam}. 
In analogy with the above example, Streltsov et al. \cite{str} have mathematically proven  that a state $\rho _{S}$ can be converted to an entangled state via incoherent operations if and only if $\rho _{S}$ is coherent (Theorem 2 of Ref. \cite{str}). Thus,  this work led to a framework for inter-conversion of quantum coherence and quantum entanglement due to which, in principle, we can use any quantifier of quantum entanglement as a measure of quantum coherence. Keeping this in mind,  a family of entanglement based coherence measures were defined  $\{ C _{E}\}$ as follows \cite{str}
\begin{equation}
C _{E} (\rho _{S})=  \lim_{d_{A} \to \infty} \{ \sup_{\wedge^{SA}}\, E^{S:A}\left( \wedge^{SA} \left[ \rho_{S}\otimes |0 \rangle \langle 0 |_{A} \right] \right) \}, 
\end{equation}
where $E$ is an arbitrary entanglement measure, and supremum is taken over all incoherent operations. 

This leads to a new perspective towards the understanding of relation between quantum coherence and quantum entanglement. However, the non-monotonic nature of the relationship between the measures of coherence is expected to remain valid for this family of measures, too, as it is well-known that the measures of entanglement are not monotones of each other \cite{adam-ent,ent_mon}. Another major criticism of the above method of calculating the coherence is that we have to undertake the maximization over infinitely many incoherent operations that are available. So the task at  hand is practically impossible and is useful only for the cases where we can get a simple analytical expression, which is possible only for a subclass of states. 

Note that the most of the proposed quantifiers \cite{coh_rev1,coh_rev2}, such as trace distance measure of coherence \cite{rana}, robustness of coherence \cite{roc}, convex roof measures of coherence \cite{yuan,winter}, geometric coherence \cite{str}, coherence monotones of entanglement \cite{str}, coherence of assistance \cite{chitambar}, are based on optimization and do not have any analytical form (except for some subset of states) or do not satisfy the property of strong monotonocity. Moreover, it has been shown that the trace distance measure of coherence \cite{rana} and robustness of coherence \cite{roc} reduce to $l1$ norm of coherence for all the single qubit and $X$ states.

\subsection{Measure of first-order coherence}

Beyond the framework of resource theory, there are some other types of coherence or asymmetry measures \cite{kagalwala,optical_coh,fang} which are also of great significance in optical coherence theory and condensed matter physics. A particular measure of first-order coherence \cite{optical_coh}, $ D = \sqrt{2\,Tr[\rho^2]-1},$ has been exploited recently for single qubit subsystems to introduce the concept of accessible coherence. For example, for a two-qubit state $\rho _{AB}$, the degree of first-order coherence of each subsystem is given by $D_{i} = \sqrt{2\,Tr[\rho_{i}^2]-1}\, \forall\, i\in \left\{A,B\right\} $, and the amount of coherence present in the system is defined as $ D^{2}= (D_{A}^{2}+D_{B}^{2})/2$. The state $\rho _{AB}$ can be transformed via a global unitary ($U$) to get state $\rho^{\prime}_{AB}= U\rho_{AB}U^{\dagger}$ such that the first-order coherence can vanish and the two subsystems become strongly correlated. On the other hand, for certain unitary operation ($U^{\prime}$), the maximum first-order coherence can be obtained, which is $ D^{2}_{\rm max} = (\lambda _{1}- \lambda _{4})^{2}+(\lambda _{2}- \lambda _{3})^{2}$, where $\lambda_i$s are the eigenvalues of the state $\rho _{AB}$ in a decreasing order \cite{optical_coh}. Further, $D^{2}_{\rm max}$ can be called the degree of available coherence, since it represents the maximum first-order coherence that can be extracted under a global unitary transformation.

\section{$X$ states}\label{X states}

$X$ states were introduced by Yu and Eberly \cite{xstate1, xstate2} in a study highlighting the finite time disentanglement of two qubits due to spontaneous emission resulting in entanglement sudden death \cite{esd}. Since then these states have become {a subject of extensive study as they} contain an important class of pure and mixed states, such as maximally entangled states (like Bell states), partially entangled and quantum correlated states (like the Werner states \cite{werner}), maximally nonlocal mixed states (MNMSs) \cite{mnms}, maximally entangled mixed states (MEMSs) \cite{munro,ishizaka,verst} as well as non-entangled (separable) states. The $X$ states are described in the computational basis $ \{ \vert 00 \rangle$, $ \vert 01 \rangle$, $ \vert 10 \rangle$,$ \vert 11 \rangle \}$ as
\begin{equation} \label{x_state}
\rho_{X} = \left(
     \begin{array}{cccc}
      \rho_{11} & 0 & 0& \rho_{14}  \\
       0&\rho_{22} &\rho_{23}& 0 \\
       0 & \rho^{*}_{23} &\rho_{33}& 0 \\
      \rho^{*}_{14} & 0 & 0 & \rho_{44} \\
     \end{array}
   \right).
\end{equation}
The positions of the non-zero elements of $\rho_X$ resemble the shape of the letter $X$, and hence the states having density matrix of this form are referred to as $X$  states. For $\rho_{X} $ to represent a physical state, we must have $ \underset{i}{\sum}\rho_{ii}=1 $, $ \rho_{22}\rho_{33}\geq |\rho_{23}|^2$, and $ \rho_{11}\rho_{44}\geq |\rho_{14}|^2$ \cite{x_dynamics}. Further, $\rho_{14} $ and $\rho_{23} $ are complex numbers, but they can always be made real and non-negative by local unitary transformations. Thus, without loss of generality, we can always start with a $\rho_{X}$ with all real and non-negative elements. The eigenstates of $\rho_{X}$ are given by
\begin{eqnarray}\label{eigenvalues}
\lambda _{1}= \frac{1}{2} \left\{(\rho_{11}+\rho_{44})+ \sqrt{(\rho_{11}-\rho_{44})^{2}+ 4|\rho_{14}|^{2}} \right\}, \\ \nonumber 
\lambda _{2}= \frac{1}{2} \left\{(\rho_{11}+\rho_{44})- \sqrt{(\rho_{11}-\rho_{44})^{2}+ 4|\rho_{14}|^{2}} \right\}, \\ \nonumber
\lambda _{3}= \frac{1}{2} \left\{(\rho_{22}+\rho_{33})+ \sqrt{(\rho_{22}-\rho_{33})^{2}+ 4|\rho_{23}|^{2}} \right\}, \\ \nonumber
\lambda _{4}= \frac{1}{2} \left\{(\rho_{22}+\rho_{33})- \sqrt{(\rho_{22}-\rho_{33})^{2}+ 4|\rho_{14}|^{2}} \right\}. 
\end{eqnarray}
As mentioned previously, $X$ states can be both separable and entangled depending upon the values of parameters describing them. It is known that $X$ states are entangled if and only if either $\rho_{22}\rho_{33} < |\rho_{14}|^{2}$ or $\rho_{11}\rho_{44} < |\rho_{23}|^{2}$ \cite{x_dynamics}, and the amount of entanglement, as measured by concurrence, is given by 
\begin{equation}
\begin{array}{lcl}
{\rm Concurrence}(\rho_{X})&=& 2 \, \max \thinspace \left\{0, |\rho_{14}|- \sqrt{\rho_{22}\rho_{33}},\right. \\
& &\left.|\rho_{23}|- \sqrt{\rho_{11}\rho_{44}} \right\}.
\end{array}
\end{equation}

Another interesting feature of  $X$ states is that they have only nonlocal coherence (i.e., the coherence pertaining to the total system). All local coherences (i.e., the coherence of the reduced density matrix of the subsystem) vanish for these states. This can be easily visualized from the reduced density matrices that can be obtained for the subsystems  $A$ and $B$ as
\begin{equation}
\rho^{A}_{X} = \left(
     \begin{array}{cc}
      \rho_{11}+\rho_{22} & 0  \\
       0 & \rho_{33}+\rho_{44}     
     \end{array}
   \right)
\end{equation}
and
\begin{equation}
    \rho^{B}_{X} = \left(
     \begin{array}{cc}
      \rho_{11}+\rho_{33} & 0  \\
       0 & \rho_{22}+\rho_{44}      
     \end{array}
   \right).
\end{equation}
For unitary time evolution, it has been observed that $\rho_{X}$ retains its $X$ form if and only if the Hamiltonian is $X$ shaped in the computational basis  \cite{x_dynamics}. Further, it is also known that this state can retain its form during the time evolution under a restricted class of open system dynamics \cite{x_dynamics}.  

Recently, numerous studies have been reported that look at the theoretical aspects of $X$ states, with specific focus on quantum correlations, as well as their production and manipulation in experimental systems. Ali et al. \cite{ali} have tried to find an analytical expression for  the quantum discord of two-qubit $X$ states, which was later found not to be valid for a few states \cite{lu}. Rau \cite{rau} has studied the algebraic characterization of $X$ states in quantum information. On the experimental side, two-qubit $X$ states can be produced in a wide variety of systems, such as optical systems \cite{xp_p1,xp_p2,xp_p3,xp_p4}, ultra cold atoms \cite{xp_uc1,xp_uc2}, and nuclear magnetic resonance \cite{xp_nmr}. The importance of $X$ states lies in the very sparse structure of the density matrix describing them, due to which they can be analyzed efficiently. Recently, Paulo et al. \cite{xh} have proved that for every two-qubit state, there is a $X$-counterpart, i.e., a corresponding two-qubit $X$ state having the same spectrum and entanglement, as measured by concurrence, negativity or relative entropy of entanglement. This universality property of  $X$ states allows us to map every two-qubit state to an equivalent $X$ state, and hence these $X$ states constitute an important resource for quantum communication and computation. The study of $X$ states is very important in understanding subtle concepts of quantum correlations, quantum coherence, and quantum entanglement as they form a very broad subset of two-qubit mixed states and incorporate most of the states that can be produced experimentally. In the next section, we will try to quantify the quantum coherence of $X$ states via different available measures of coherence and try to find relations between them.

\section{Relations between the resource theoretic measures of quantum coherence}\label{our-res}

\begin{figure}[!htb]
\includegraphics[width=8.6 cm]{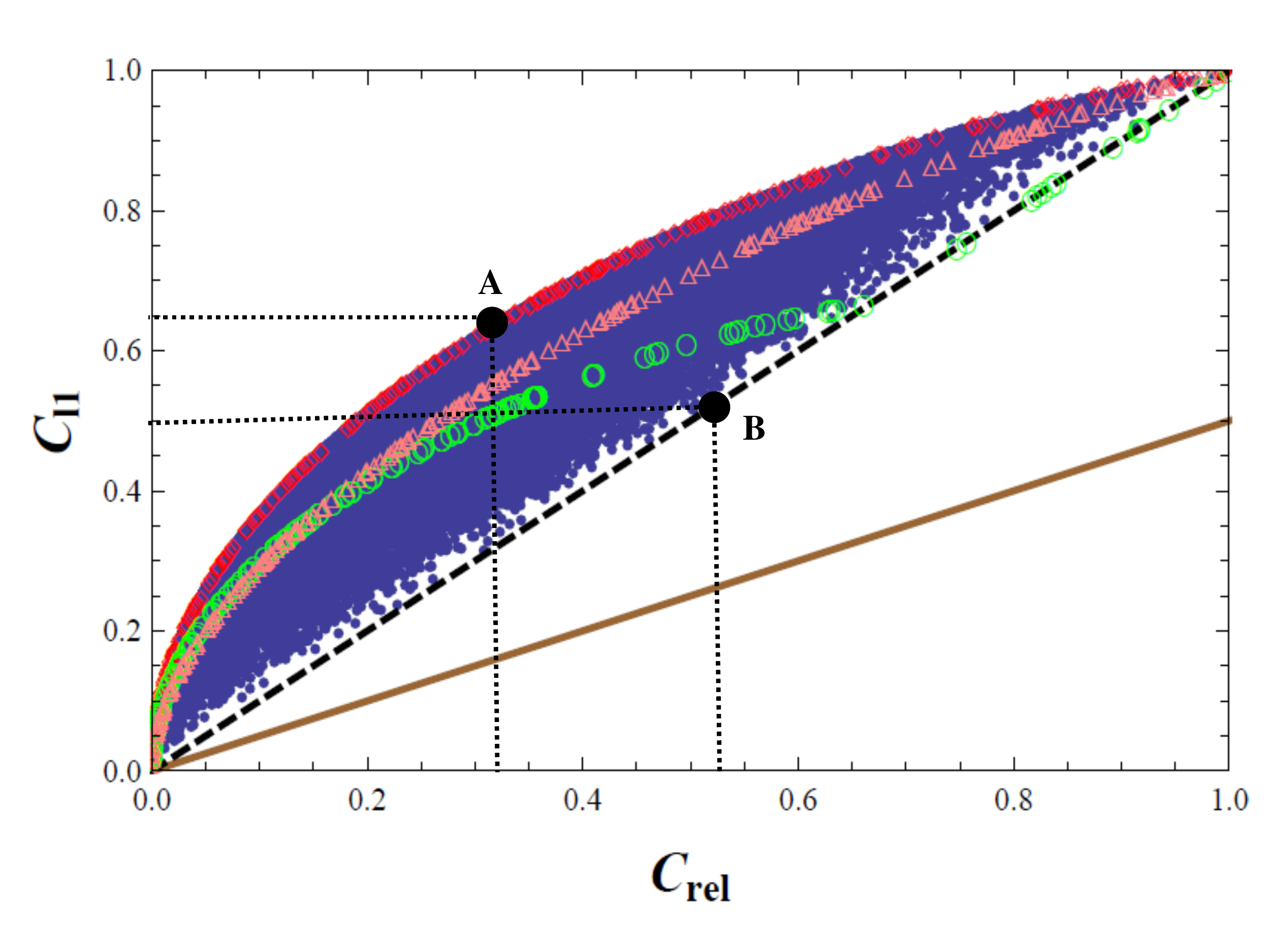}
\caption{\label{fig:epsart1} (Color online) The blue points represent scatter plots for $X$ states. The red squares represent the same for MNMSs, and the green circles correspond the same for MEMSs. The pink triangles are obtained for the Werner states. The black (dashed) and brown (smooth) curves represent the line with slope 1 and 0.5 respectively. The black (dashed) line also represents the results obtained for state $\rho_L$. All the quantities shown in this plot and the rest of the figures in the present work are dimensionless.}
\label{fig1}
\end{figure}

We have  randomly prepared $10^5$ $X$ states and have quantified the amount of coherence present in these states using the quantitative measures of coherence described in Section \ref{measures}. Specifically, for each of the randomly prepared states we have computed  the relative entropy of coherence, $l1$ norm of coherence, coherence via skew information, and first-order coherence. The obtained results are plotted to reveal relationships between various measures of coherence. To begin with, in Fig. \ref{fig1}, we provide a scatter plot of random $X$ states with coherence measured by relative entropy of coherence on the abscissa and $l1$ norm of coherence on the ordinate. The relative entropy of coherence (\ref{rel_ent}) for the $X$ states can be expressed as 
\begin{equation}\label{rel_ent_x}
C_{\rm{} rel}(\rho _{X})= \sum_{i} \lambda _{i} \log_{2}(\lambda_{i})- \sum_{i} \rho _{ii} \log_{2}(\rho _{ii})  , 
\end{equation}
where $\lambda _ {i} $s are the eigenvalues of the $X$ states given by Eq. (\ref{eigenvalues}), while $\rho _{ii} $ represent the diagonal values of the $X$ state (\ref{x_state}). Similarly, the amount of coherence of $X$ states (\ref{x_state}) measured by $l1$ norm of coherence (\ref{l1}) can be expressed as
\begin{equation}\label{cl1_ent_x}
C_{l1}(\rho_{X})= 2(|\rho_{14}|+|\rho_{23}|).
\end{equation}
We can clearly see from Fig. \ref{fig1} that these two quantum coherence quantifiers are not monotone of each other. To illustrate this point specifically, we have marked two points on the plot as A and B which correspond to two different $X$ states $\rho_A$ and $\rho_B$, respectively. Clearly, as far as the relative entropy of coherence is concerned $\rho_B$ has more coherence than $\rho_A$. However, the opposite is observed if we measure coherence using $l1$ norm of coherence. Thus, we cannot conclude whether  $\rho_A$ possesses more coherence than $\rho_B$ or not. This situation is analogous to the case of measures of nonclassicality \cite{adam,adam-non}, entanglement \cite{adam-ent}, steering \cite{adam-st}, Bell nonlocality \cite{adam-bi}, and non-Markovinaity \cite{adam-nm}, where non-monotonic natures of different measures have already been observed.

The MNMSs \cite{mnms} form a subclass of $X$ states and are described as 
\begin{equation}
\rho_{\rm{} MNMS} = \left(
     \begin{array}{cccc}
      \frac{1}{2} & 0 & 0& \frac{\epsilon}{2}  \\
       0 & 0 & 0 & 0 \\
       0 & 0 & 0 & 0 \\
      \frac{\epsilon}{2} & 0 & 0 & \frac{1}{2} \\
     \end{array}
   \right).
\end{equation}
For each value of $\epsilon\in\left\{0,1 \right\}$, the state $\rho_{\rm{} MNMS}$ is a Bell diagonal state and represents the state that produces a maximal violation of the CHSH inequality \cite{chsh}.  In the scatter plot of relative coherence for the random $X$ states,  measured by relative entropy of coherence and $l1$ norm of coherence, we find that the MNMSs form the upper boundary as represented by {the red squares} in Fig. \ref{fig1}. Clearly, this shows that for a given value of quantum coherence measured by the relative entropy of coherence, the MNMSs of $X$ type  have the maximum quantum coherence as measured by $l1$ norm of coherence. Further, we can note that  these states having maximum quantum coherence also maximally violate the CHSH inequality.

The Werner states \cite{werner}, which are described as a statistical mixture of a maximally entangled state and a maximally mixed state can be written as
\begin{eqnarray}
\rho_{\rm{} W} = \epsilon | \Phi^{+}\rangle \langle \Phi^{+} | + \frac{1- \epsilon}{4} I_{2}\otimes I_{2},
\end{eqnarray}
where $ I_{2}$ is {the} identity matrix, and $| \Phi^{+}\rangle = \frac{1}{\sqrt{2}}\{ |00 \rangle + | 11 \rangle \} $. Note that the Werner state is also a subset of $X$ states as it can be written as 
\begin{equation}
\rho_{\rm{} W} = \left(
     \begin{array}{cccc}
      \frac{1+\epsilon}{4} & 0 & 0& \frac{\epsilon}{2}  \\
       0 &\frac{1-\epsilon}{4} & 0 & 0 \\
       0 & 0 & \frac{1-\epsilon}{4} & 0 \\
      \frac{\epsilon}{2} & 0 & 0 & \frac{1+\epsilon}{4} \\
     \end{array}
   \right).
\end{equation}
Depending on the value of $\epsilon$, a Werner state {is} entangled if $\epsilon > \frac{1}{3}$ and separable otherwise \cite{werner2}. For $\frac{1}{3} < \epsilon < \frac{1}{\sqrt{2}} $, the Werner states  are found to be entangled, but such states do not violate any  Bell’s inequality. The Werner states, sometimes referred to as decoherence-free states, have special significance in quantum information processing applications where there is a need to combat decoherence in noisy channels \cite{nc}. We can clearly see from Fig. \ref{fig1} that the amount of coherence measured by $l1$ norm of coherence for a Werner state (pink triangles) is always less than that for a MNMS (red squares) even though both  the states may have the same amount of coherence as measured by the relative entropy of coherence. 

Let us further consider the case of MEMSs \cite{munro}. These represent a class of states for which no additional entanglement can be produced by global unitary operations. These states  are a  generalization of the class of Bell states to mixed states and are known to have the  highest degree of entanglement for a given purity of a state. While Bell states are {known to be} maximally entangled two-qubit {pure} states, all other maximally entangled states can be represented as
\begin{equation}
\rho_{\rm MEMS} = \left(
     \begin{array}{cccc}
      g (\epsilon) & 0 & 0& \frac{\epsilon}{2}  \\
       0 & 1- 2g (\epsilon) & 0 & 0 \\
       0 & 0 & 0 & 0 \\
      \frac{\epsilon}{2} & 0 & 0 & g (\epsilon) \\
     \end{array}
   \right),
\end{equation}
where
\begin{equation}
g (\epsilon)= \begin{cases}
  \frac{\epsilon}{2}, & \text{if $\epsilon \geq \frac{2}{3}$ } \\
  \frac{1}{3}, & \text{$\epsilon < \frac{2}{3}$}
\end{cases}.
\end{equation}
We can clearly see from Fig. \ref{fig1} that the MEMSs  (represented by the green circles) have lesser quantum coherence as measured by $l1$ norm of coherence than that of the MNMSs and Werner states for the same amount of quantum coherence as measured by relative entropy of coherence.

Interestingly, Rana et al. \cite{rana} have proved that for any $d$ dimensional mixed state $ C_{l1} (\rho) \geq  \frac{C_{\rm{} rel}(\rho)}{\log_{2} d}$ and conjectured that $ C_{l1} (\rho) \geq C_{\rm{} rel} (\rho)$ for all states. The smooth brown line in  Fig. \ref{fig1} represents a straight line with slope $\frac{1}{2}$, while the black dashed curve represents the straight line with slope $1$. We can clearly see from Fig. \ref{fig1} that Rana et al.'s conjecture described by inequality (\ref{conj}) is clearly valid for the case of $X$ states. Let us now focus on the states which would satisfy $ C_{l1} (\rho_{X}) = C_{\rm{} rel} (\rho_{X})$, i.e., the $X$ states that have the same amount of quantum coherence as measured by $l1$ norm of coherence and relative entropy of coherence. These states are given by 
\begin{equation}
\rho_{L} = \left(
     \begin{array}{cccc}
      \frac{\epsilon}{2} & 0 & 0& \frac{\epsilon}{2}  \\
       0 & 1- \epsilon & 0 & 0 \\
       0 & 0 & 0 & 0 \\
      \frac{\epsilon}{2} & 0 & 0 & \frac{\epsilon}{2} \\
     \end{array}
   \right)
\end{equation}
for $0 \leq \epsilon\leq 1$. {The amount} of coherence present in these  states as measured by two different measures of coherence has  been illustrated by the black dashed line in Fig. \ref{fig1}. We can clearly see that $C_{l1}(\rho_{L})= \epsilon $, and the eigenvalues of $\rho_{L}$ are $0,0,$ $ \epsilon$, and $1-\epsilon$. Using the eigenvalues of $\rho_{L}$ in Eq. (\ref{rel_ent}), the relative entropy of coherence evaluates to the  value of $C_{\rm{} rel}(\rho_{L})= \epsilon $, which is same as that computed using the $l1$ norm of coherence. We can see that the  state $\rho_{L}$ is similar to the MEMS for $\epsilon \geq \frac{2}{3}$ and retains its form for the whole range ($0 \leq \epsilon\leq 1$).

In summary, we have seen that the two well-known measures of quantum coherence, namely the relative entropy of coherence and the  $l1$ norm of coherence, are not monotone of each other, and we have analytical expressions for the states forming the lower and upper bounds of the scatter plots (Fig. \ref{fig1}) for the class of $X$ states. 

To further stress on the non-monotonic nature of the measures of quantum coherence, we now consider another measure of coherence, which is referred to as trace distance measure of coherence \cite{rana} and is defined as $C_{tr}(\rho)= \min ||\rho - \delta ||_{1} $, where $\delta$ belongs to the set of incoherent states $I$. The problem with this measure is that it does not satisfy the property of strong monotonicity for all the states. Rana et al. \cite{rana} have shown that for the case of single qubit and $X$ states, the trace distance measure of coherence reduces to just the $l1$ norm of coherence and hence is a valid measure of quantum coherence for these states. Thus, Fig. \ref{fig1} also establishes the fact that the trace distance measure of coherence and the relative entropy of coherence are not monotone of each other. Similarly, it is known that the quantum coherence as measured by the robustness of coherence \cite{roc} reduces to the $l1$ norm of coherence for $X$ states. Thus, Fig. \ref{fig1} also establishes that robustness of coherence  and the relative entropy of coherence are not monotone of each other.  

\begin{figure}[!htb]
\includegraphics[width=8.6 cm]{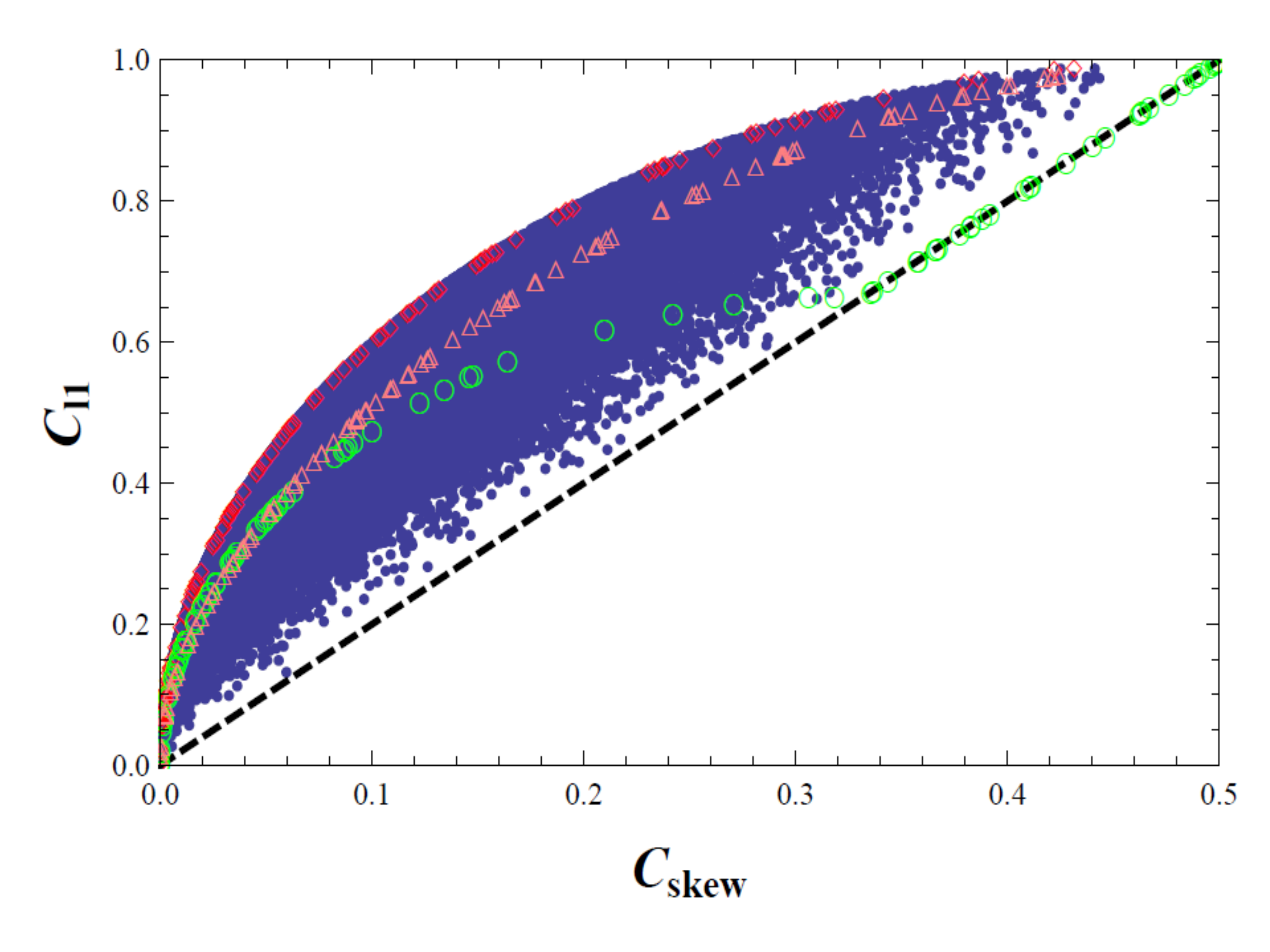}
\caption{\label{fig:epsart1} (Color online) The blue points represent scatter plots of $X$ states for $l1$ norm of coherence versus the quantum coherence via skew information. The red squares, green circles, and pink triangles represent the same for MNMSs, MEMSs, and Werner states, respectively. Further, the black (dashed) line represents the results obtained for state $\rho_L$.}
\label{fig2}
\end{figure}

\begin{figure}[!htb]
\includegraphics[width=8.6 cm]{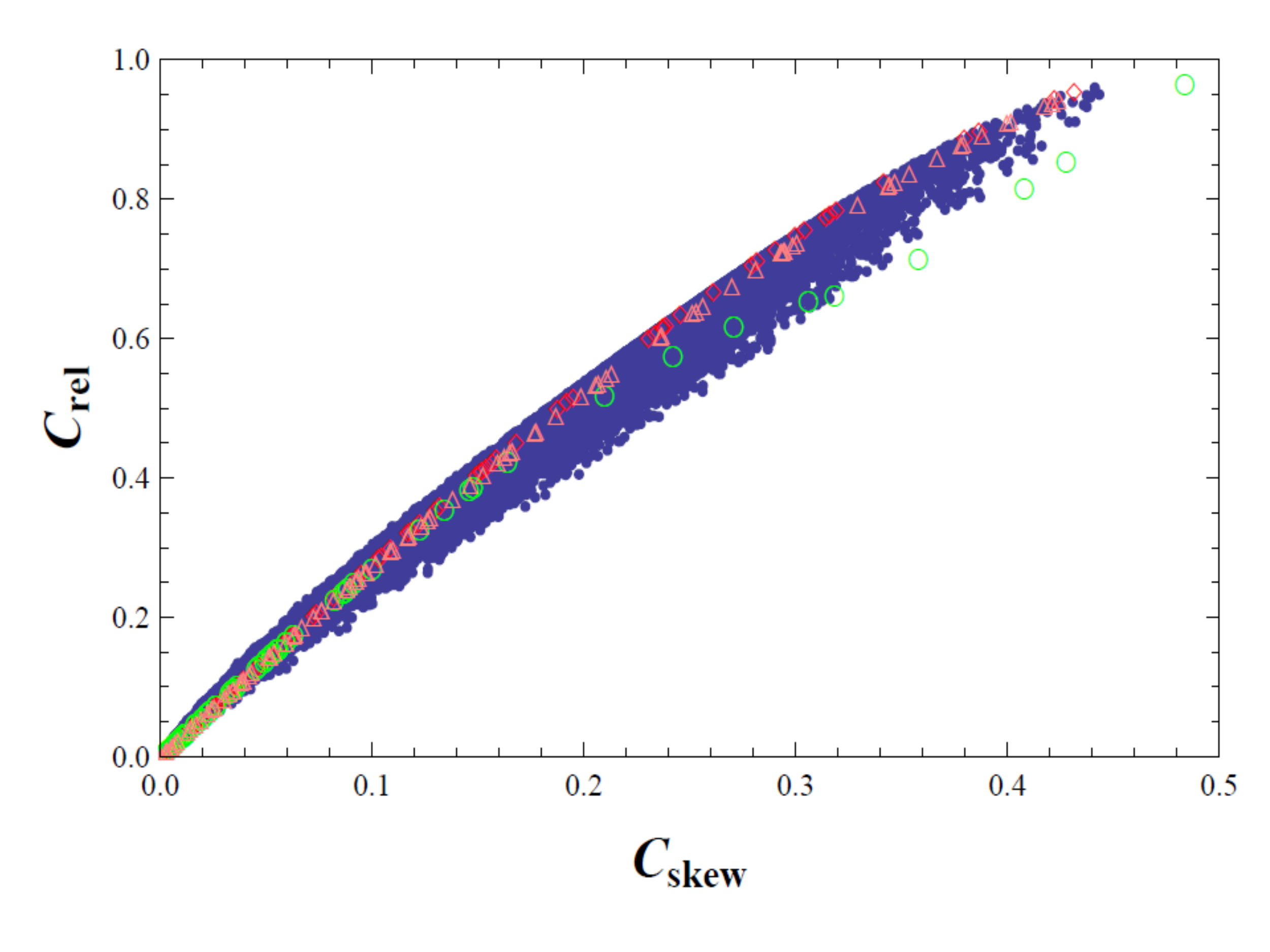}
\caption{\label{fig:epsart1} (Color online) The blue points represent scatter plots of $X$ states for relative entropy of coherence versus the quantum coherence via skew information. The red squares, green circles, and pink triangles represent the same for MNMSs, MEMSs, and Werner states, respectively.}
\label{fig3}
\end{figure}

In Sec. \ref{measures}, we have already defined the skew information based measure of coherence (\ref{skew}). Here, we would try to compute this measure of coherence for $X$ states and check whether this measure is monotonic with the other measures, i.e.,  relative entropy of coherence and $l1$ norm of coherence. Figure \ref{fig2} represents the scatter plot of $l1$ norm of coherence and quantum coherence via skew information for $X$ states. We can clearly see that these two measures are also not monotone of each other. Interestingly, just like the variation of $l1$ norm of coherence with relative entropy of coherence, MNMSs ($\rho_{\rm{} MNMS}$) are found to form the upper boundary of the scatter plot, and the state ($\rho_{L}$) forms the lower boundary, while the MEMSs and Werner states are observed to lie between them. Therefore, the quantifiers of quantum coherence, namely $l1$ norm of coherence and quantum coherence via skew information, are also not monotonic of each other and will not follow each other with respect to the ordering of the states. Figure \ref{fig3} represents the scatter plot of the relative entropy of coherence versus quantum coherence via skew information, and we can clearly see that these two measures of coherence are also not monotone of each other. In the light of all our results and observations, we can conclude that the different popular quantifiers currently being used  to measure the quantum coherence are not equivalent.  This would imply an ambiguity with respect to the ordering of the states as can be seen from the plots in Figs. \ref{fig1}-\ref{fig3}. Further, we have already mentioned that the non-monotonic nature of the measures of coherence shown in this paper  would remain valid for  entanglement-based measures of coherence too, as it is well-known that the measures of entanglement are also not monotone of each other \cite{ent_mon}. 

\begin{figure}[!htb]
\includegraphics[width=8.6 cm]{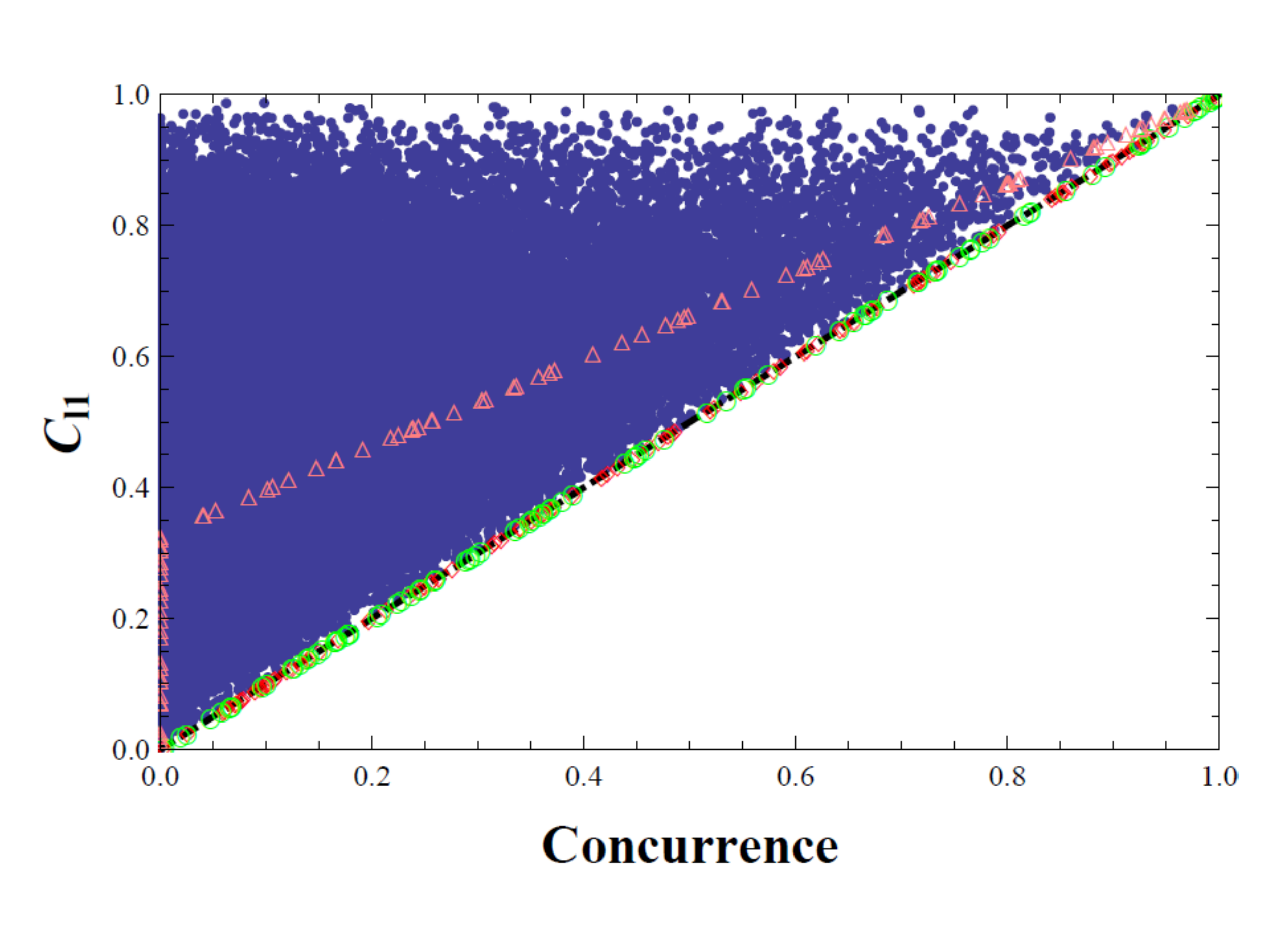}
\caption{\label{fig:epsart1} (Color online)  The blue points represent scatter plots of $l1$ norm of coherence versus quantum entanglement measured by concurrence for the $X$ states. The red squares, green circles, and pink triangles represent the same for MNMSs, MEMSs, and Werner states, respectively. }
\label{fig4}
\end{figure}

Let us look further at the possible relation between the amount of entanglement as measured by concurrence and the amount of coherence as measured by $l1$ norm of coherence. For the $X$ states under consideration, we can see that $C_{l1}(\rho_{X}) = 2 \left(|\rho_{14}|+|\rho_{23}|\right),$  while the concurrence is given by $ {\rm Concurrence}(\rho_{X})= 2 \, \max \thinspace \{0, |\rho_{14}|- \sqrt{\rho_{22}\rho_{33}}, |\rho_{23}|- \sqrt{\rho_{11}\rho_{44}} \}$. Thus, we can clearly see that $C_{l1}(\rho_{X})\geq {\rm Concurrence}(\rho_{X})$. We further analyze the state for which $C_{l1}(\rho_{X})= {\rm Concurrence}(\rho_{X}),$ i.e., the amount of coherence as measured by the $l1$ norm of coherence is equal to the amount of entanglement as measured by concurrence. We can see from Fig. \ref{fig4} that MNMSs  and MEMSs form the lower boundary of the scatter plot for concurrence and $l1$ norm of coherence, respectively. Further, we know that Werner states are separable if $ \epsilon < \frac{1}{3}$ and entangled otherwise. This fact is also reflected in Fig. \ref{fig4}.  {Note that here, all the coherence measures, which are basis dependent, are obtained in the computational basis. There is a simple reason behind it- the $X$ states are defined in the computational basis only. However, in principle, we can measure coherence using other bases, too.  If we change the basis and compute the coherence using different measures, it is expected that the analytical form of the states forming the upper and lower boundary would change.  Specifically, we may visualize this point, by noting that MNMSs are Bell diagonal states, and we have already shown that these states form the upper boundary in our Fig. \ref{fig1}. Therefore, if we choose the Bell basis as our incoherent basis, then MNMSs will not form the upper boundary. Here, we restrict ourselves from exploring more in this direction as the study on the computational basis alone provides us answer to the question that this paper aims to address. Keeping this in mind, we now proceed to describe a new measure of coherence in the  following section (first-order coherence), which is a basis independent measure like concurrence. }

\begin{figure}[!htb]
\includegraphics[width=8.6 cm]{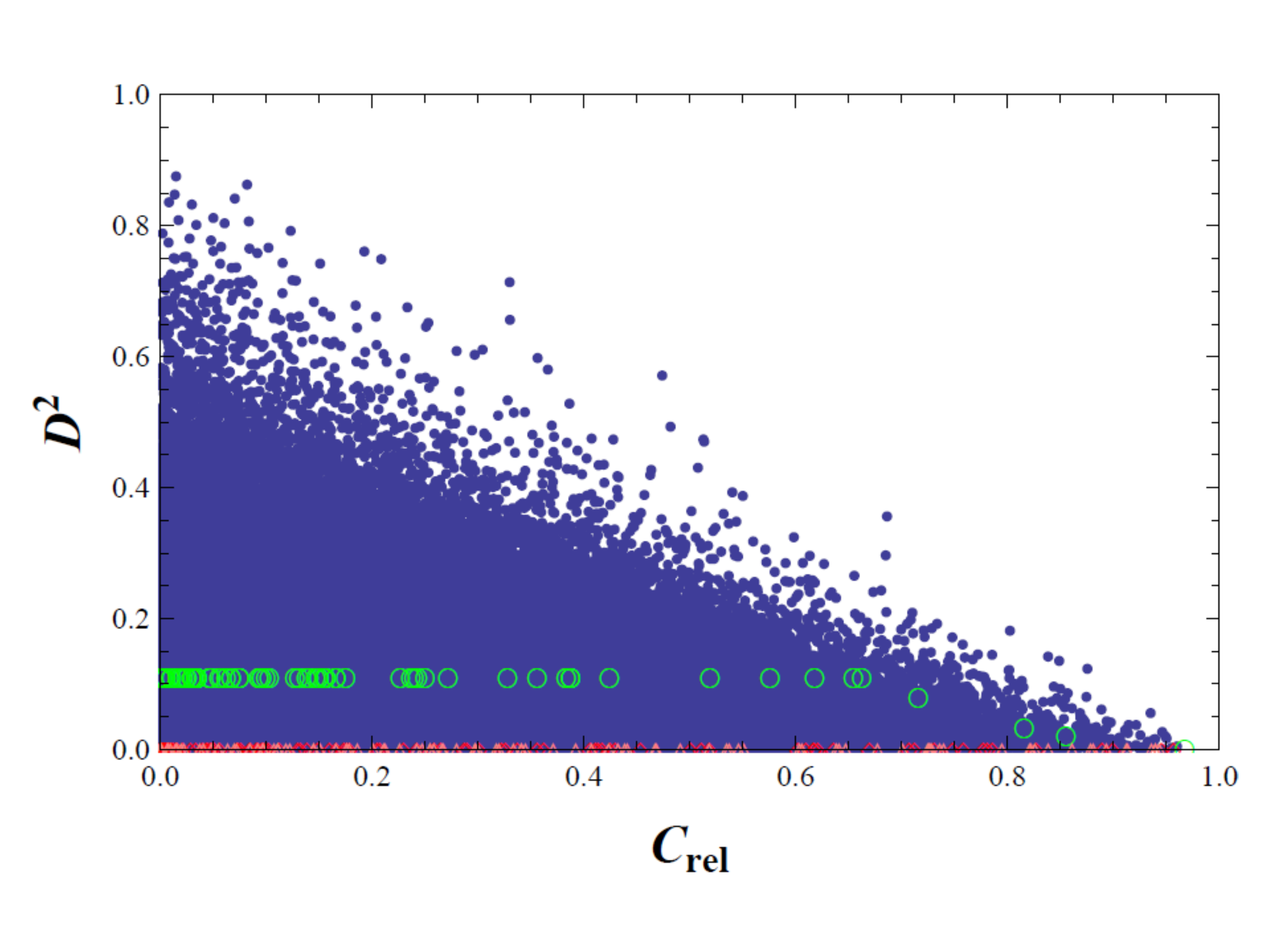}
\caption{\label{fig:epsart1} (Color online)  The blue points represent scatter plots of first-order coherence ($D^{2}$) versus relative entropy of coherence. The red squares, green circles, and pink triangles represent the same for MNMSs, MEMSs, and Werner states, respectively.}
\label{fig5}
\end{figure}

\begin{figure}[!htb]
\includegraphics[width=8.6 cm]{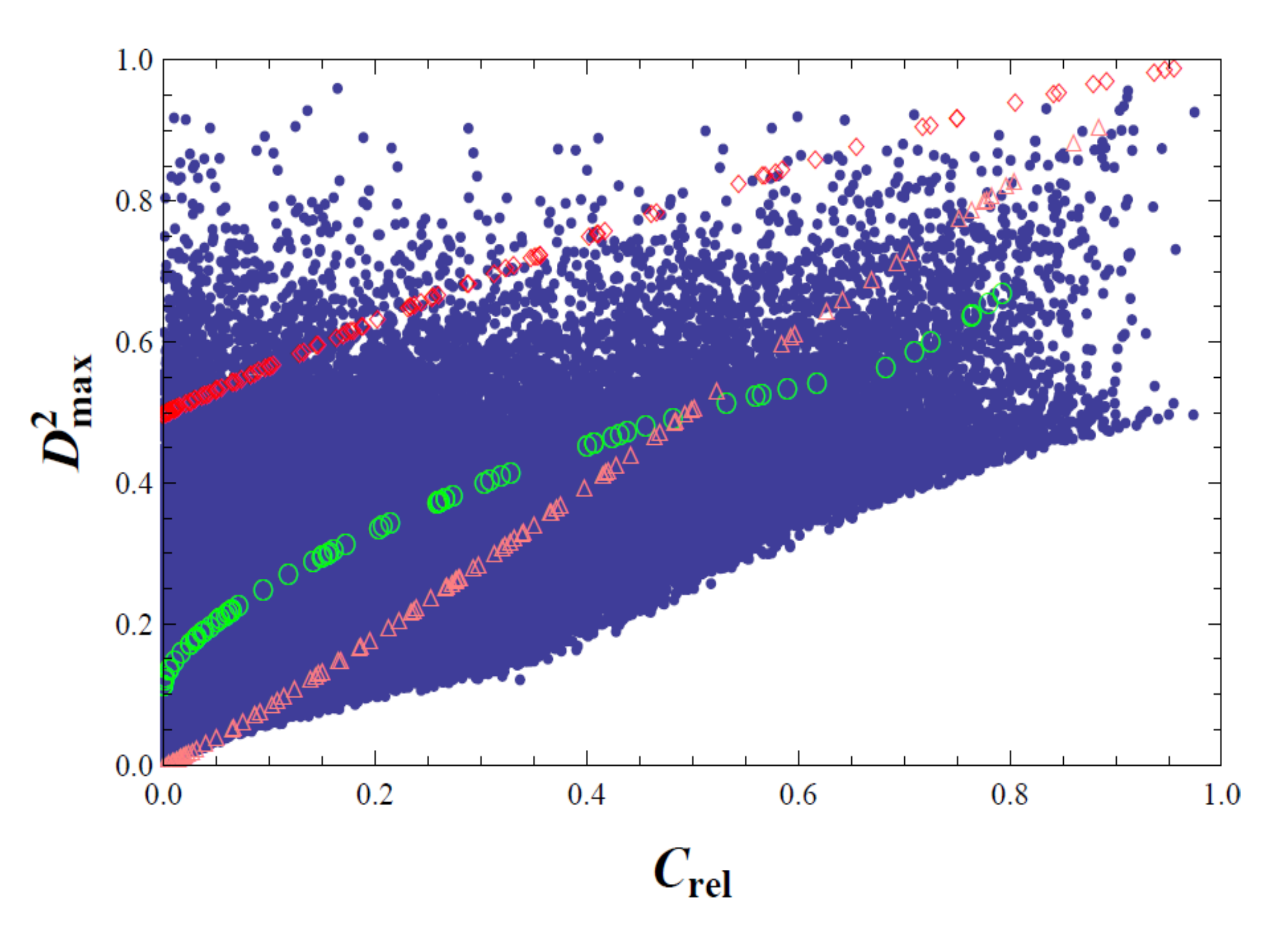}
\caption{\label{fig:epsart1} (Color online) The blue points represent scatter plots of hidden coherence ($ D^{2}_{\rm max}$) versus relative entropy of coherence. The red squares, green circles, and pink triangles represent the same for MNMSs, MEMSs, and Werner states, respectively.}
\label{fig6}
\end{figure}

\begin{figure}[!htb]
\includegraphics[width=8.6 cm]{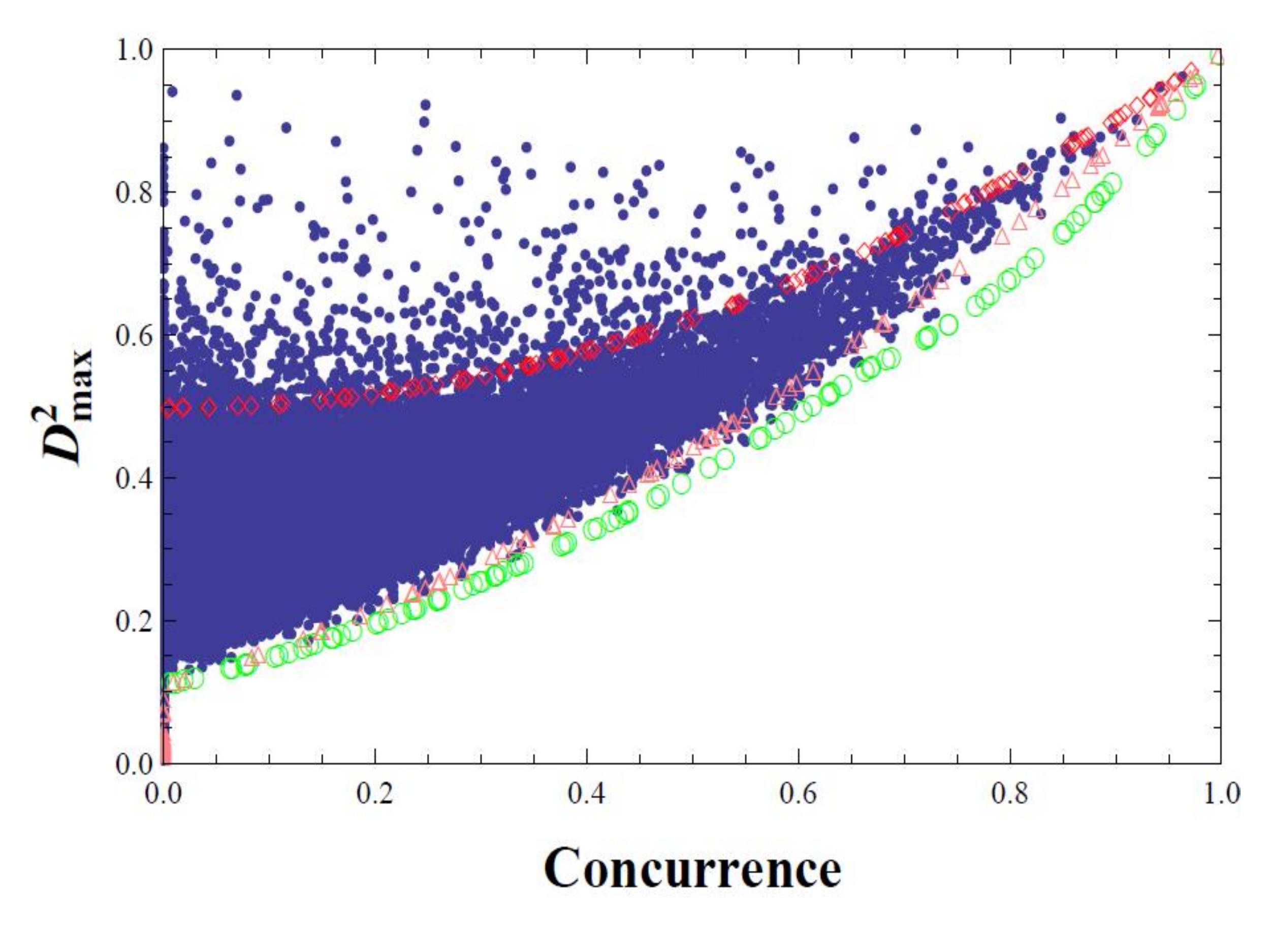}
\caption{\label{fig:epsart1} (Color online) The blue points represent scatter plots of hidden coherence ($ D^{2}_{\rm max}$) versus concurrence. The red squares, green circles, and pink triangles represent the same for MNMSs, MEMSs, and Werner states, respectively.}
\label{fig7}
\end{figure}

\section{First-order coherence}\label{Op-coh}

Let us further analyze, how the first-order coherence \cite{optical_coh} and the maximum first-order coherence vary for  $X$ states.  This coherence measure is based on the purity of the subsystems which constitute the bipartite state. From Figs. \ref{fig5} and \ref{fig6}, we can see that there is no clear relation between the first-order coherence and the measures of coherence as described by resource theory of coherence, such as $l1$ norm of coherence, relative entropy of coherence, and coherence using the skew information. This was expected as first-order coherence was introduced with an altogether different motivation, and it does not follow  Baumgratz et al.'s criteria. Moreover, it is related to the purity of the individual subsystems of which the combined bipartite system is composed.  In Section \ref{measures}, we have already mentioned that the amount of hidden coherence (degree of available coherence) is known as the maximum first-order coherence. Consequently, this measure of coherence is   a basis independent measure in contrast to the basis dependent measures of coherence present in the resource theory of quantum coherence discussed above. Svozil{\'\i}k et al. \cite{optical_coh} have shown a possible trade off between the amount of hidden coherence present in the system and the amount of violation of CHSH inequality \cite{chsh}.

Specifically, Figs. \ref{fig5} and \ref{fig6} illustrate the scatter plot of $ D^{2}$ and $ D^{2}_{\rm max} $ versus $C_{\rm rel}$ for different $X$ states under consideration.  We can clearly see from Fig. \ref{fig5} that the amount of first-order coherence, i.e., $D^{2}$ is zero for  the case of MNMS (red squares) and Werner (pink triangles) states, while it is nonzero for most of the MEMSs (green circles). Further, the amount of hidden coherence $ D^{2}_{\rm max}$, i.e., the coherence available after the unitary transformation for $X$ states shows that of {all the} subclasses of $X$ states considered in the study, MNMSs (red squares) have the maximum amount of  hidden coherence $ D^{2}_{\rm max}$ with respect to the relative entropy of coherence. Therefore, we can see that the maximum first-order coherence is also not monotonic with the measures of quantum coherence studied here. Specifically, it is illustrated in Fig. \ref{fig6}  that the maximum first-order coherence is not monotonic with the relative entropy of coherence. Further, it is checked that the maximum of first-order coherence is not monotonic with $l1$ norm of coherence and skew information based measure of coherence, too. However, corresponding plots are not shown here. We have also studied the relation between first-order coherence $( D^{2})$ and the amount of entanglement as measured by concurrence for different $X$ states. However, we have not included the corresponding plot as it is found to be  similar to the scatter plot of $ D^{2}$  versus  $C_{\rm rel}$ (Fig. \ref{fig5}).

Interestingly, both $ D^{2}_{\rm max}$ and  concurrence are known to be basis independent quantities. We found this fact motivating enough to explore the relationship between these two basis independent quantities. Figure \ref{fig7} illustrates the scatter plot of $ D^{2}_{\rm max}$ with respect to the concurrence of  the $X$ states under consideration. We can clearly see from Fig. \ref{fig7} that the MEMSs (green circles) form the lower boundary of the plots. We can see that for zero value of concurrence (i.e., for the separable state), MEMSs (green circles) have a value of $ D^{2}_{\rm max}=0.1$, while for MNMSs (red squares)  we obtain $D^{2}_{\rm max}=0.5$. However, there are separable $X$ states (including the separable subclass of Werner states) which can be observed to have smaller values of hidden coherence than that of MEMSs. As the amount of concurrence starts to increase, the difference in the amount of hidden coherence $ D^{2}_{\rm max}$ for the MNMSs (red squares) and the MEMSs (green circles) starts decreasing and becomes equal to zero when  the amount of concurrence = 1 (i.e., for a  maximally entangled $X$ state). In all cases, for the same amount of entanglement as measured by concurrence, the amount of hidden coherence for the MNMSs (red squares) is always greater that that of the MEMSs (green circles).

The study of first-order coherence has provided a kind of completeness to the present study, but we could not find any concrete relation between first-order coherence and other measures of coherence studied here. However, efforts have already been made to relate the resource theories of coherence and the interferometric visibility (cf. Bera et al. \cite{bera} and Bagan et al. \cite{bagan}) by using $l1$ norm of coherence to measure the visibility in a multi-slit experiment \cite{tania, biswas, sandeep}.

\section{Conclusions}\label{con}

This paper aimed to answer the question: Can we compare the quantum coherence present in two states? A detailed analysis revealed that the answer is no. This is so because the analysis performed using the $X$ states and the measures of coherence (relative entropy of coherence, $l1$ norm of coherence, skew information based measure of  {coherence,} robustness of coherence, trace distance norm of coherence, and first-order  coherence) has proved that the measures of coherence studied here are not monotone of each other. {This feature (non-monotonic nature of the measures) is not only present in the nonclassicality measures reported before \cite{adam,adam-non} but also in measures of coherence (studied here). Similar results  have also been observed in the context of measures of entanglement \cite{adam-ent}, steering \cite{adam-st}, Bell nonlocality \cite{adam-bi}, non-Markovianity \cite{adam-nm}, etc. Specifically, in our analysis, we see that in all these cases  some of the investigated measures are not monotones of each other. Further, our analysis reveals that for a given value of quantum coherence measured by the relative entropy of coherence, the MNMSs of $X$ type  have the maximum quantum coherence as measured by $l1$ norm of coherence. In addition, we observe that the amount of coherence measured by $l1$ norm of coherence for a Werner state is found to be always less than that for a MNMS even when they possess an equal amount of coherence as measured by the relative entropy of coherence. We have illustrated our main observations in graphs (Figs. \ref{fig1}-\ref{fig7}). Further, we have also found analytical expressions for the states forming the upper and lower bounds of the scatter plots of the $l1$ norm of coherence and the relative entropy of coherence for $X$ states. It is interesting that the same behavior was observed between the $l1$ norm of coherence and skew information based measure of coherence, i.e., the boundary states are observed to be the same as that in the previous case. However, no such relation between the relative entropy of coherence and skew information based measure of coherence have been found as the states with both higher and lower values of $C_{\rm rel}$ were observed for the same amount of $C_{\rm skew}$. Further, we have analyzed the results for the case of first-order coherence to check for any relation between these two completely different types of coherence measures. However, first-order coherence being connected with the purity of reduced states has no direct relation with any measure used in the resource theory of quantum coherence.  Also, considering its close analogy with entanglement measured using concurrence, we have studied the relation between first-order coherence and concurrence to reveal that they are not related. However, the maximum of first-order coherence is found to be more related to concurrence as both are basis independent quantities. Note that neither first-order, nor hidden coherence show monotonic behavior with concurrence. Also, during the present study, we have restricted to computational basis as our incoherent basis as $X$ states are defined in this basis only and a different choice of incoherent basis would have revealed the same results with different boundary states.

We have not only shown the conjecture in Ref. \cite{rana} to satisfy in the present study (at least for $X$ states), using some of the existing discrete results on the quantum coherence measured for $X$ states, we have  extended our results to a large class of measures of quantum coherence and have shown that they too are not monotone of each other. Specifically, the trace distance and robustness of coherence have the same value as that of $l1$ norm of coherence for $X$ states and are thus non-monotonic with relative entropy of coherence and skew information based measure of coherence as well.
Finally, we conclude that the quantum coherence measures studied here are not monotone of each other. Probably, at a deeper level they capture different manifestations of nonclassicality.  Consequently, there is no way to circumvent  the difficulty associated with the comparison of the amount of coherence between two states. Further, the relationship of coherence with different measures of nonclassicality, entanglement and other measures of quantum correlations, such as discord, is still an open problem and this work is expected to provide some deeper understanding of these facets of nonclassicality and their mutual relationship.

\section*{Acknowledgment}
AP thanks  Department  of  Science and  Technology (DST), India  for  the support provided
through the project number EMR/2015/000393. SM thanks Guru Gobind Singh Indraprastha university for STRF.  KT thanks the project LO1305 of the Ministry of Education, Youth and Sports of the Czech Republic for support. AP and KT also thank A. Miranowicz, P. Panigrahi and J. Perina Jr. for some fruitful discussion and their interest in this work.

\end{document}